%% file: main.tex
\documentclass[a4paper]{article}
\usepackage{Odyssey2022}
\usepackage{epsfig,amssymb,amsmath, multirow,cite,subfigure}

\usepackage{color}

\ninept

\title{Tackling Spoofing-Aware Speaker Verification with Multi-Model Fusion}

\makeatletter
\def\name#1{\gdef\@name{#1\\}}
\makeatother
\name{ {\em Haibin Wu$^{12}$, Jiawen Kang$^{3}$, Lingwei Meng$^{3}$, Yang Zhang$^{4}$, Xixin Wu$^{3}$,} \\ 
{\em Zhiyong Wu$^{4}$, Hung-yi Lee$^{1}$, Helen Meng$^{23}$}}

\address{$^1$ Graduate Institute of Communication Engineering, National Taiwan University \\
$^2$ Centre for Perceptual and Interactive Intelligence, The Chinese University of Hong Kong \\
$^3$ Human-Computer Communications Laboratory, The Chinese University of Hong Kong \\
$^4$ Shenzhen International Graduate School, Tsinghua University 
}
\begin{document}

\maketitle

\input{0-abstract-keywords}

\input{1-introduction}

\input{2-method}

\input{3-setup}

\input{4-experiment}

\input{5-conclusion}

\input{6-acknowledge}

\input{main.bbl}
\bibliographystyle{IEEEbib}
% \bibliography{Odyssey2022_BibEntries}

\end{document}

%% file: 0-abstract-keywords.tex
\begin{abstract}
Recent years have witnessed the extraordinary development of automatic speaker verification (ASV).
However, previous works show that state-of-the-art ASV models are seriously vulnerable to voice spoofing attacks, and the recently proposed high-performance spoofing countermeasure (CM) models only focus solely on the standalone anti-spoofing tasks, and ignore the subsequent speaker verification process.
How to integrate the CM and ASV together remains an open question.
A spoofing aware speaker verification (SASV) challenge has recently taken place with the argument that better performance can be delivered when both CM and ASV subsystems are optimized jointly.
Under the challenge's scenario, the integrated systems proposed by the participants are required to reject both impostor speakers and spoofing attacks from target speakers, which intuitively and effectively matches the expectation of a reliable, spoofing-robust ASV system. 
This work focuses on fusion-based SASV solutions and propose a multi-model fusion framework to leverage the power of multiple state-of-the-art ASV and CM models. 
The proposed framework vastly improves the SASV-EER from 8.75\% to be 1.17\%, which is 86\% relative improvement compared to the best baseline system in the SASV challenge.
\end{abstract}

%% file: 1-introduction.tex
\section{Introduction}
\label{sec:intro}
Automatic speaker verification (ASV) is a technology to verify claimed speakers by the given speech segments. Resorting to the power of deep neural networks (DNNs) in representation learning, speaker embedding-based ASV approaches have achieved state-of-the-art performances on several benchmark datasets \cite{snyder2018x,variani2014deep,okabe2018attentive,nagrani2017voxceleb, chung2018voxceleb2, fan2020cn, li2022cn,zhang2022conformer,desplanques2020ecapa,zeinali2019but,li2017deep,li2015improved}. 
Despite the impressive performance, current ASV models are vulnerable to spoofing attacks, which include audio replay, text-to-speech (TTS) and voice conversion (VC) \cite{wu2015asvspoof,kinnunen2017asvspoof,todisco2019asvspoof, yamagishi2021asvspoof,Yi2022ADD,wang2021comparative}, back-door attacks \cite{zhai2021backdoor} and related emerged adversarial attacks \cite{kreuk2018fooling, das2020attacker}. Various methods have been proposed to tackle adversarial attacks \cite{wang2019adversarial,li2020investigating,zhang2020adversarial,wu2021adversarialasv,wu2021improving,joshi2021adversarial,wu2021voting,wu2021spotting}.

At the same time, led by a series of challenges \cite{wu2015asvspoof, kinnunen2017asvspoof, todisco2019asvspoof, yamagishi2021asvspoof}, many research efforts have investigated countermeasure (CM), also known as anti-spoofing, to defend and detect spoofing attacks. 
The current methods utilize end-to-end neural network structures \cite{monteiro2019development, monteiro2020generalized} to distinguish between spoofing speech from bona fide speech by either the raw speech signal or hand-crafted features. 
Recently, CM systems have been significantly improved by a series of works \cite{monteiro2019development, monteiro2020generalized,zhang2021one, wu2022partial,chen2020generalization, jung2021aasist, tak2021end,li2021replay}.
CM models are vulnerable to adversarial attacks \cite{liu2019adversarial}, and some efforts are dedicated to address such attacks \cite{wu2020defense_2,wu2020defense}.

While ASV and CM systems have been successfully developed as individual tasks, limited efforts have been devoted to their integration.
% While ASV and CM systems are successfully developed in respective tasks, limited efforts have been put into integrating the two standalone systems together. 
Jung et al. \cite{jung2022sasv} recently held a spoofing aware speaker verification (SASV) challenge, which is a special session in ISCA INTERSPEECH 2022, with the argument that better performance can be delivered when CM and ASV subsystems are both optimized.
They released official protocols and baselines based on the ASVspoof 2019 dataset \cite{yamagishi2019asvspoof}.
Compared with previous anti-spoofing challenges that measure CM systems based on fixed ASV models, this challenge considers CM and ASV together as an integrated system and uses a variant of classic equal error rate (SASV-EER) as a primary assessment metric.
Under this metric, the integrated systems are required to reject both impostors speakers and spoofing attacks from target speakers, which intuitively and effectively matches the expectation of a reliable, spoofing-robust ASV system. 
% Toward such SASV systems, only limited works are conducted. These solutions can be concluded as (1) fusion-based solutions \cite{sizov2015joint, gomez2020joint, todisco2018integrated}, which fuses the scores or embeddings from separate ASV and CM systems leveraged by ensemble techniques. 
% And (2) integrated single model solutions that achieve the goal by a single model with multi-task learning \cite{li2019multi, li2020joint, zhao2022multi}.

In this work, we focus on fusion-based SASV solutions and propose a multi-model fusion framework to leverage the power of multiple state-of-the-art ASV and CM models. 
The ASV and CM models we covered include MFA-Conformer \cite{zhang2022conformer}, Resnet34 \cite{zeinali2019but}, ECAPA-TDNN \cite{desplanques2020ecapa} and AASIST \cite{jung2021aasist}, AASIST-L, RawGAT-ST \cite{tak2021end}.
Based on the proposed framework, we study combinations of different ASV and CM models.
They vastly improve the SASV-EER from 8.75\% to 1.17\%, which achieves 86\% relative improvement compared to the best baseline system in the SASV challenge.
% Furthermore, (placeholder for any experimental finding)
% Our main contributions are as shown below:
% \begin{itemize}
%     \item We proposed
%     \item We further 
% \end{itemize}

% Recent years have witnessed the extraordinary development of automatic speaker verification (ASV).
% However, previous works show that the state-of-the-art ASV are serious subject replayed speech, synthesised speech \cite{wu2015asvspoof,kinnunen2017asvspoof,todisco2019asvspoof, yamagishi2021asvspoof, Yi2022ADD, wu2015spoofing,wu2014study,kamble2020advances,das2020assessing, chenglong2021global,wang2021comparative, wu2022partial}.
% ASVspoof series \cite{wu2015asvspoof,kinnunen2017asvspoof,todisco2019asvspoof, yamagishi2021asvspoof} arouse keen interest from both the academia and industry to address the replayed speech and synthesised speech, and variety of high performance anti-spoofing models have been proposed.
% However, 

% This paper is organized as follows:
% related works are summarized in Section 2.
% Section 3 presents the proposed method, multi-model feature ensemble for spoofing-aware speaker verification. 
% Section 4 illustrates the experimental setups, followed by the experimental results and discussions in Section 5.
% The conclusions are presented in Section 6.

\begin{figure*}[ht]
  \centering
  \centerline{\includegraphics[width=1.0\linewidth]{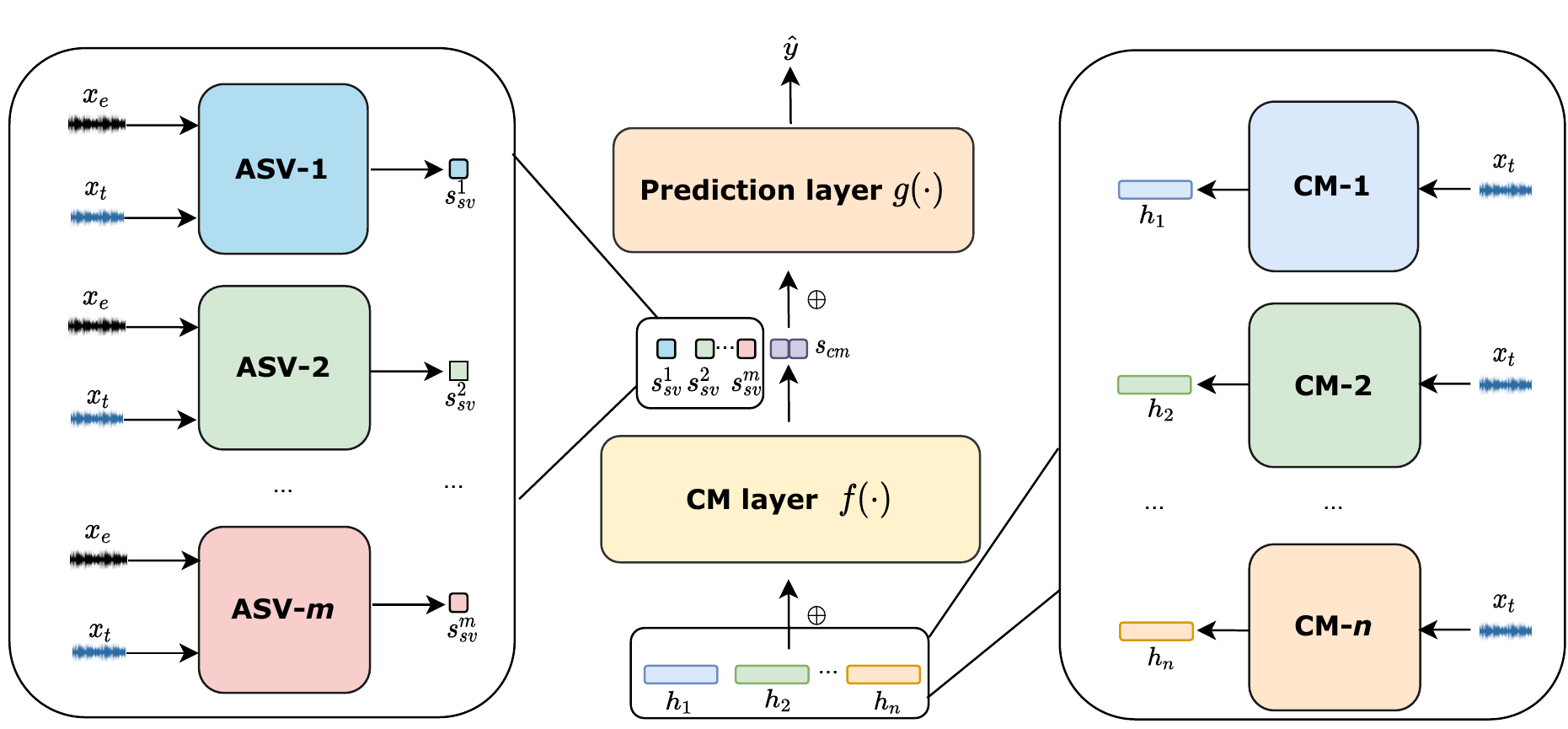}}
  \vspace{5pt}
  \caption{The proposed multi-model fusion framework.}
  \label{fig:method}
\end{figure*}

\section{Related work}

\subsection{Automatic speaker verification}
Automatic speaker verification aims to determine whether a given utterance belongs to the claimed speaker. 
Most modern ASV approaches are based on speaker embeddings, i.e., representing speech segments into fix-length identity vectors. Speaker verification can be achieved by measuring the distance between different speaker embeddings. 
These methods train the models to distinguish speakers from the variable-length speech signal. Then an embedding of the bottleneck layer is extracted as a speaker-dependent vector to identify speaker information.
Following the speaker embedding scheme, current ASV systems mainly involve two steps: 
% firstly, the speaker embeddings of enrollment and test speech are extracted, representing the identification of target speakers and verified speech segment, respectively.
the speaker embeddings are extracted from the enrollment and test speech to respectively represent the identification of the target speaker and verified speech segment.
Then, a metric is used to score the test speech based on enrollment speech, where the metric can be cosine distance or more advanced statistical models like PLDA \cite{ioffe2006probabilistic}.
After many years' development, many technologies and typologies have been introduced to further improve speaker embedding models, including better structures \cite{chung2018voxceleb2, desplanques2020ecapa, zeinali2019but}, training schemes \cite{gao2019improving, zhou2019deep} and pooling approaches \cite{ xie2019utterance, chen2019tied}.
% As a result, the current ASV models have achieved a equal error rate (EER) of less than 1\%.
% Traditionally, this embedding is derived from probabilistic models, e.g., the i-vector model \cite{dehak2010front}. 
% More recently, DNN-based models have become the mainstream solutions.
% In current, best-performed methods are based on x-vector \cite{snyder2018x} system with different improvements in terms of better structures \cite{chung2018voxceleb2, desplanques2020ecapa, zeinali2019but}, training scheme \cite{gao2019improving, zhou2019deep} and pooling approaches \cite{ xie2019utterance, chen2019tied}.

\subsection{Countermeasures}
Countermeasure approaches for ASV systems are developed to detect the spoofing samples generated by attack techniques, including text-to-speech, voice conversion, and audio replay. 
Conventional CM models use a Gaussian Mixture Model (GMM) classifier as back-end with different input features \cite{patel2015combining, sahidullah2015comparison, todisco2016new}.
In \cite{zhang2017investigation}, a combination of CNN and RNN models is proposed as a DNN-based solution. 
\cite{monteiro2020generalized} introduced Resnet in this task, which is further improved by \cite{chen2020generalization} using large margin cosine loss and frequency masking data augmentation. 
\cite{tak2021end} was first to introduce RawNet that directly uses raw speech waveform as input.
\cite{tak2021graph} introduced graph attention networks (GAT) with the ability of modeling the relationships between neighbouring sub-bands or segments of spectrum features.
Based on GAT, a series of extensions are proposed in \cite{tak2021end, jung2021aasist} using both raw audio and spectrum features.
Besides, a one-class learning method was proposed in  \cite{zhang2021one} for better generalization ability.
Such previous efforts presented a variety of anti-spoofing models with high performance.
% Thanks to previous efforts, a variety of anti-spoofing models with high performance have been proposed.

% \subsection{Spoofing aware speaker verification system (SASV)}
% As mentioned above, the conception of SASV is firstly proposed in SASV Challenge 2022. 
% There are limited works trying to optimize ASV and CM models. 
% For fusion-based solutions, \cite{gomez2020joint} fuses the embeddings of ASV and CM model by an integrated neural network.
% \cite{sizov2015joint} proposed a method to model synthesis-channel subspace and perform SASV in i-vector space.
% Apart from embedding-based fusion, an alternative approach is scoring fusion.
% Apart from such hidden embedding-based approaches, many works also considered score fusion, which is done by Gaussian back-end fusion \cite{todisco2018integrated}, cascaded/parallel fusion framework \cite{sahidullah2016integrated} and optimizing a differentiable detection cost function using reinforcement learning \cite{kanervisto2021optimizing}.

% For integrated single model solutions, an common idea is to obtain a joint embedding representing both ASV and CM information by multi-task learning. \cite{li2019multi} achieved this by a modified triple loss, and \cite{zhao2022multi} used sequential residual convolutional blocks with Max-Feature-Map activations.

\label{sec:related}

%% file: 2-method.tex
\section{Methodology}
\label{sec:method}

\subsection{SASV systems}
There are two main categories of spoofing-aware speaker verification (SASV) systems: multi-task learning based systems and fusion based systems.
The multi-task learning systems are trained by both the anti-spoofing loss and speaker verification loss.
The speaker verification task and anti-spoofing task share the same input features and the embedding backbone layers.
However, such a system requires both speaker labels and anti-spoofing labels for training.
Only the ASVspoof dataset \cite{wang2020asvspoof} can fulfill such requirements.
The ASVspoof dataset contains only a small number of speakers, which will induce the trained multi-task learning system to overfit to only a small number of speakers, resulting limited generalization capacity to real-world applications \cite{jung2022sasv}.
% Another disadvantage of such systems is that the learning objectives of speaker verification systems and anti-spoofing systems are different: the anti-spoofing models rely on the device information to detect the spoofing utterances, 
Another disadvantage of such systems is that the learning objectives of anti-spoofing and speaker verification systems are different -- anti-spoofing models rely on the device information to detect the spoofing utterances, yet the speaker verification systems tend to remove such information which is incongruent with speaker discrimination \cite{shim2020integrated}.

The other kind of SASV system is the fusion-based systems, which may involve embedding-level or score-level fusion.
% They can do both embedding-level and score-level fusion.
Such systems can rely on state-of-the-art speaker verification models trained on large-scale datasets \cite{nagrani2017voxceleb,chung2018voxceleb2} and state-of-the-art spoofing countermeasure models.

Thus, the SASV challenge adopts the second type of systems as baselines.
We will introduce the details of such systems in the next subsection.

\subsection{Baseline systems}
The SASV Challenge 2022 provides two baselines. 
The organizers provided one pre-trained speaker verification model, ECAPA-TDNN, and one anti-spoofing model, AASIST.
Both models attain state-of-the-art performance for their own tasks.

Baseline1 is a score-level fusion method. For a given trial, which consists of an enrollment utterance and a testing utterance, ECAPA-TDNN will be used to obtain the enrollment embedding and the testing embedding from the enrollment and testing utterances.
% The Baseline1 is a score-level fusion method.
% Given a trial, which is consist of an enroll utterance and a testing utterance.
% They will use the ECAPA-TDNN to get the enroll embedding and testing embedding from the enroll and testing utterance.
Then the cosine score of the enroll embedding and testing embedding is calculated.
The anti-spoofing score is also obtained from the spoofing countermeasure (CM), AASIST, given the testing utterance as the input. 
Subsequent addition of the SV and CM scores produces the final score, which indicates whether such a trail is target trial or not.
% They also will get the anti-spoofing score from the spoofing countermeasure (CM), AASIST given the testing utterance as the input.
% Then they do addition of the SV score and CM score to get the final score, indicating whether such a trail is target trial or not.

% The Baseline2 is an embedding-level fusion method.
% They will use the ECAPA-TDNN to get the enroll embedding and testing embedding, and use AASIST to get the countermeasure embedding.
Baseline2 is an embedding-level fusion method. ECAPA-TDNN is used to obtain the enrollment embedding and testing embedding, and AASIST is used to obtain the countermeasure embedding.
These three embeddings are concatenated and fed to a deep neural network to predict whether the trial belongs to the target or not.
% Then they do concatenation to the three pieces of embedding.
% Finally a deep neural network is adopted to take the concatenation as input, and predict whether the trial belongs to target or not.

\subsection{Proposed systems}
The baselines of the SASV Challenge do achieve better performance, compared with using speaker verification models or anti-spoofing models alone, as shown in Table~\ref{tab:all-eer}.
However, Baseline1 does not consider that the score scales of ASV score and CM score are in different ranges, and Baseline2 does not consider fine-grained fusion.
Thus they cannot take good advantage of the state-of-the-art ASV and anti-spoofing models in boosting their performances.

This motivates us to propose an ensemble model for the SASV task.
We will first report the details of the framework followed by rational analysis.
The proposed framework is shown in Figure~\ref{fig:method}.
ASV-1, ASV-2, ..., ASV-$m$ denote $m$ ASV pre-trained models.
CM-1, CM-2, ..., CM-$n$ denote $n$ countermeasure models.
$x_{e}$ and $x_{t}$ are the enrollment and testing utterances, respectively. 
Given a trial composed of an enrollment and testing utterance, $m$ cosine scores, $s_{sv}^{1}, s_{sv}^{2},...,s_{sv}^{m}$ are calculated by $m$ pre-trained speaker verification models. 
$\oplus$ denotes the concatenation procedure. 
$f$ and $g$ are the trainable countermeasure (CM) layer and prediction layer, respectively.
Given the testing utterance, $n$ testing embeddings, $h_{1}, h_{2},...,h_{n}$ are extracted by $n$ pre-trained anti-spoofing models, respectively. 
Then the $n$ testing embeddings are firstly concatenated, and then fed into the countermeasure layer to get the countermeasure score $s_{cm}$.
We use the cross-entropy loss on $s_{cm}$ to guide the CM layer learn to distinguish between the target trials and other trials:
\begin{equation}
    L_{cm}= -log \frac{exp(s^{l}_{cm})}{\sum_{j=0}^{1} exp(s^{j}_{cm})},
    \label{eq:cm-loass}
\end{equation}
where $l \in \{0,1\}$ is the label, indicating whether the trail is target or not.
Then we concatenate $s_{cm}$ and $s_{sv}^{1}, s_{sv}^{2},...,s_{sv}^{m}$, and then input them to the prediction layer to get the final prediction $\hat{y}$
Cross-entropy is derived to train the entire model:
\begin{equation}
    L_{pr}= -log \frac{exp(\hat{y}_{l})}{\sum_{j=0}^{1} exp(\hat{y}_{j})},
    \label{eq:pr-loass}
\end{equation}
where $l \in \{0,1\}$ is the label, denoting whether the input trail is target or not.
To be specific, 1 denotes the non-spoofed target trial.
The final loss is:
\begin{equation}
    L=L_{cm}+L_{pr}.
    \label{eq:final-loss}
\end{equation}

% The previous years have witnessed the development of speaker verification models and current state-of-the-art speaker verification models are trained on a very large dataset.
Previous years have witnessed the development of state-of-the-art speaker verification models that are trained on very large datasets.
Well-trained ASV models attain very good performance for in-domain data as shown in Table~\ref{tab:asv eer}, and have good generalization and reliability for out-of-domain data as shown in Table~\ref{tab:all-eer}.
As a result, we directly use the scores provided by the ASV for fusion.
Anti-spoofing models detect artifacts located in specific spectrum segments or temporal segments \cite{jung2021aasist}, and different anti-spoofing models focus on different kinds of artifact.
Thus we propose to ensemble different CM embeddings from a group of countermeasure models.
The results shown in Table~\ref{tab:all-eer} and Figure~\ref{fig:hist} show the effectiveness of the proposed method.

%% file: 3-setup.tex
\section{Experimental setup}

\subsection{Dataset}
The SASV Challenge restricts the training data for ASV to the VoxCeleb2 development set, and the training data for the anti-spoofing model to ASVspoof 2019 training set.
We strictly follow the constraints released by the Challenge and more details can be found in \cite{jung2022sasv}.

Released with Automatic Speaker Verification Spoofing And Countermeasures Challenge 2019, the ASVspoof 2019 dataset aims to determine whether the advances in text-to-speech and voice conversion technology pose increased threat to automatic speaker verification and the reliability of spoofing countermeasures \cite{wang2020asvspoof}. 
It aligns the anti-spoofing task closely with the ASV task, and sparks the community's interest in the SASV Challenge.
The ASVspoof 2019 dataset consists of bona fide speech trails collected from the VCTK corpus \cite{yamagishi2019vctk}, and spoofing attack trials synthesised by text-to-speech, voice conversion, or replay. 
The speakers, with their corresponding bona fide and spoofing trails, are partitioned into three subsets, namely the training (20 speakers with 2,580 bona fide trails and 22,800 spoofing trails), development (20 speakers with 2,458 bona fide trails and 22,296 spoofing trails) and evaluation sets (67 speakers with 7,355 bona fide trails and 63,882 spoofing trails). Moreover, the spoofing attack algorithms used in the evaluation set are different from those in training and development set, which is intended to promote the generalization performance of the developed systems. Based on the ASVspoof 2019 dataset, the SASV Challenge provided the development and evaluation protocols which list target, non-target, and spoof trials, and for each them, multiple enrolment utterances exist. 

For ASV training, VoxCeleb2 \cite{chung2018voxceleb2} is a benchmark dataset for speaker verification, containing audio clips collected from YouTube. There are totally 1,092,009 utterances from 5,994 speakers in the VoxCeleb2 development set.

\subsection{Evaluation metrics}
Three kinds of EERs, SV-EER SPF-EER and SASV-EER, are used by the SASV Challenge, and SASV-EER on the ASVspoof 2019 evaluation set is adopted as the main evaluation metric.
Equal error rate (EER)  is the overall error rate when a threshold is set such that the false accept rate equals the false reject rate. The lower the EER, the better the discrimination capacity the model. Widely used in binary classification problems, it is a more objective and comprehensive metric than accuracy, especially on data with unbalanced classes.

In this challenge, EERs are used to evaluate the performance of the models on three sub-tasks: ASV, anti-spoofing, and SASV \cite{jung2022sasv}. The definition of the EERs are showed in Table~\ref{tab:metric}. For all three sub-tasks, trails from the target speaker are treated as positive samples, while the definition of negative samples is different. For the SASV sub-task, both the non-target speakers' bona fide speech trails and spoofing trails are considered negative samples. For ASV sub-task, only the non-target speakers' bona fide speech trails are considered as negative samples. For anti-spoofing sub-task, only spoofing trails aimed at target speaker are considered as negative samples, which is different from other general anti-spoofing tasks.

\begin{table}[htb]
\centering
\renewcommand\arraystretch{1.2}
\caption{ Description of EERs. "+" denotes positive samples, and "-" denotes negative samples. Blank denotes samples not considerated in the calculation of the EER.}
\vspace{5pt}
\begin{tabular}{lcccc}
\hline
\hline
Metrics  & Target & Non-target & Spoof             \\ 
\hline
SASV-EER   & + & - & -    \\    
SV-EER    & + & - &    \\    
SPF-EER      & + &   & - \\    
\hline
\hline
\label{tab:metric}
\end{tabular}
\end{table}

\subsection{Implementation setup}

\subsubsection{ASV}
\label{subsec:asv}
Following the requirements of the SASV challenge, we only use the development set of VoxCeleb2 as the training data.
% \textcolor{red}{VoxCeleb2 dataset is a benchmark dataset for speaker verification, containing audio clips collected from the YouTube.
% There are totally 1,092,009 utterances from 5,994 speakers in VoxCeleb2 development set.}
During training, we randomly extract 3-second segment of each segments and use the on-the-fly feature extraction procedure.
The input features for training are 80-dimensional Mel spectrogram with the window length of 25 ms and the hop size of 10 ms.
No data augmentation and voice activity detection are applied.

For training the speaker verification backbone models, we adopt three well-known structures, namely, MFA-Conformer \cite{zhang2022conformer}, ECAPA-TDNN \cite{desplanques2020ecapa} and Resnet34 \cite{he2016deep}.
For the ECAPA-TDNN, we use the pre-trained model provided by the SASV Challenge.
MFA-Conformer and Resnet34 are trained by the additive margin Softmax (AM-Softmax) loss \cite{wang2018cosface} with the margin as 0.2 and the scaling factor as 30.
The speaker embedding dimensions of all three models are 192.
Adam optimizer with an initial learning rate of 0.001 is adopted and the learning rate decreases by 50\% every 4 epoch.
Also the weight decay is set as 1e-7 to avoid overfitting.

Then we evaluate the three models using equal error rate (EER) on the VoxCeleb1-O testing set \cite{nagrani2017voxceleb}.
Cosine scoring is implemented and the results are shown in Table~\ref{tab:asv eer}.
As we can see, the MFA-Conformer backbone achieves the best EER of 0.91\%, which outperforms the ECAPA-TDNN provided by the SASV challenge. 

\begin{table}[htb]
\centering
\renewcommand\arraystretch{1.2}
\caption{ASV performance on VoxCeleb1-O testing set.}
\vspace{5pt}
\begin{tabular}{lcccc}
\hline
\hline
Model  & Frontend & EER(\%)             \\ 
\hline
MFA-Conformer   & Mel spectrogram &  \textbf{0.91}    \\    
Resnet34    & Mel spectrogram &   1.50    \\    
ECAPA-TDNN  & Mel spectrogram &  0.96    \\    
\hline
\hline
\label{tab:asv eer}
\end{tabular}
\end{table}

\subsubsection{Anti-spoofing}
\label{subsec:cm}
For the anti-spoofing model, we adopt AASIST \cite{jung2021aasist}, AASIST-L, and RawGAT-ST \cite{tak2021end}. 
AASIST-L is the light version of AASIST. 
The AASIST pre-trained model is provided by the SASV challenge.
The training data for the three models is from the logical access of ASVspoof 2019 \cite{wang2020asvspoof}, which is composed of spoofing audios generated by text-to-speech and voice conversion.
% \textcolor{red}{The dataset consists of three parts: train, development and evaluation. 
% The train and development sets are composed of the same spoofing attack method (A1-A6).
% While the evaluation set contains other unseen attacks (A7-A19).}
The evaluation metrics for the anti-spoofing model on the evaluation set are min-tDCF and EER. 
As we can see in Table~\ref{tab:cm eer}, AASIST is with the best performance, namely the lowest min-tDCF and EER.

\begin{table}[htb]
\centering
\renewcommand\arraystretch{1.2}
\caption{Anti-spoofing model performance on ASVspoof 2019 evaluation set.}
\vspace{5pt}
\begin{tabular}{lcccc}
\hline
\hline
Model  & Frontend &min-tDCF & EER(\%)             \\ 
\hline
RawGAT-ST   & Raw waveform &  0.0335 & 1.06    \\    
AASIST-L    & Raw waveform &  0.0309 & 0.99   \\    
AASIST      & Raw waveform &  \textbf{0.0275} & \textbf{0.83}    \\    
\hline
\hline
\label{tab:cm eer}
\end{tabular}
\end{table}

\subsubsection{Fusion method}
We adopt the models in Section~\ref{subsec:asv} and Section~\ref{subsec:cm} as the ASV models and CM models, respectively, as shown in Figure~\ref{fig:method}.
For the CM layer $f$ in Figure~\ref{fig:method},we use three fully connected layers with hidden dimensions 256, 128 and 64 respectively.
For the prediction layer $g$, we use one linear fully connected layer.
We train our model using Adam optimizer with an initial learning rate as 0.0001.
The batch size is 32 and the epoch number is set as 200.
We selected the best model based on the development set of the ASVspoof 2019 dataset.

%% file: 4-experiment.tex
\section{Experimental results and analysis}
\label{sec:expt}

\begin{table*}[ht]
\centering
\renewcommand\arraystretch{1.2}
\setlength\tabcolsep{13pt}
\caption{Performance of all systems on the ASVspoof 2019 development and evaluation dataset. Three evaluation metrics, namely SV-EER, SPF-EER and SASV-EER are illustrated.}
% \vspace{5pt}
\label{tab:all-eer}
\begin{tabular}{lccccccc}
\hline
\hline

 & \multirow{2}{*}{\textbf{Model}} &  \multicolumn{2}{c}{\textbf{SV-EER}} & \multicolumn{2}{c}{\textbf{SPF-EER}} & \multicolumn{2}{c}{\textbf{SASV-EER}} \\ 
  &  & dev & eval & dev & eval & dev  & eval \\ 
\hline
(A1) & ECAPA-TDNN   & 1.64   & 1.86  & 20.28  & 30.75 & 17.37 & 23.84     \\
(A2) & MFA-Conformer  & 1.61   & 1.38  & 19.94  & 30.22 & 16.91 & 23.28
 \\
(A3) & Resnet34  & 1.68   & 1.08  & 17.40  & 29.76 & 14.62 & 22.69
 \\
\hline
(B1) & AASIST   & 46.01   & 49.24  & 0.07  & 0.67 & 15.86 & 24.38
 \\
(B2) & AASIST-L  & 48.30   & 49.04  & 0.13  & 0.84 & 15.72 & 24.81
 \\
(B3) & RawGAT-ST & 51.25   & 49.24  & 0.34  & 0.96 & 15.96 & 24.85
 \\
\hline
(C1) & Baseline1 \cite{jung2022sasv}   & 14.89   & 35.10  & 6.94  & 0.50 & 2.09 & 19.15
 \\
(C2) & Baseline2 \cite{jung2022sasv}  & 14.38   & 16.01  & 0.01  & 1.23 & 5.41 & 8.75
 \\ 
\hline
(D1) & MFA-Conformer\_AASIST   & 1.48   & 1.47  & 0.2  & 1.08 & 0.88 & 1.35
 \\
(D2) & ECAPA-TDNN\_AASIST  & 1.48 & 1.58  & 0.2  & 1.06 & 0.99 & 1.42
 \\
(D3) & Resnet34\_AASIST & 1.49  & 1.02  & 0.20 & 1.53 & 0.88 & 1.32
 \\
(D4) & MFA-Conformer\_AASIST-L & 1.68   & 1.68  & 0.15 & 2.03 & 1.21 & 1.83
 \\
(D5) & ECAPA-TDNN\_AASIST-L  & 1.61   & 1.62 & 0.19  & 2.18 & 1.10 & 1.92
 \\
(D6) & Resnet34\_AASIST-L & 1.68  & 1.69  & 0.18  & 2.01 & 1.15 & 1.84
 \\
(D7) & MFA-Conformer\_RawGAT-ST   & 2.16   & 2.09 & 0.35  & 0.78 & 1.55 & 1.82
 \\
(D8) & ECAPA-TDNN\_RawGAT-ST  & 2.56  & 2.17 & 0.04  & 0.78 & 1.81 & 1.94
 \\
(D9) & Resnet34\_RawGAT-ST & 2.09   & 1.97 & 0.40  & 0.79 & 1.55 & 1.69
 \\
\hline
(E1) & MFA-Conformer\_CM-ALL   & 1.91  & 1.66 & 0.20 & 0.64 & 1.01 & 1.30
 \\
(E2) & ECAPA-TDNN\_CM-ALL  & 1.39   & 1.73 & 0.20  & 0.74 & 0.81 & 1.40
 \\
(E3) & Resnet34\_CM-ALL & 1.28 & 1.12  & 0.26 & 1.43 & 0.74 & 1.32
 \\
 \hline
(F1) & SV-ALL\_AASIST   & 1.42  & 1.30  & 0.27  & 1.61 & 0.81 & 1.41
 \\
(F2) & SV-ALL\_AASIST-L  & 1.42  & 1.33  & 0.47 & 3.99 & 0.88 & 2.95
 \\
(F3) & SV-ALL\_RawGAT-ST & 1.82  & 1.64  & 0.4  & 0.82 & 1.28 & 1.39
 \\
 \hline
(G1) & SV-ALL\_CM-ALL & 1.27   & 1.20  & 0.20 & 1.15 & 0.81 & 1.17
 \\
\hline
\hline
\end{tabular}
\end{table*}

In this section, we will first show the experimental results, followed by the experimental observations and analysis.

\begin{figure*}[ht]
\centering
 
\subfigure[MFA-Conformer]{
\centering
\includegraphics[width=0.32\linewidth]{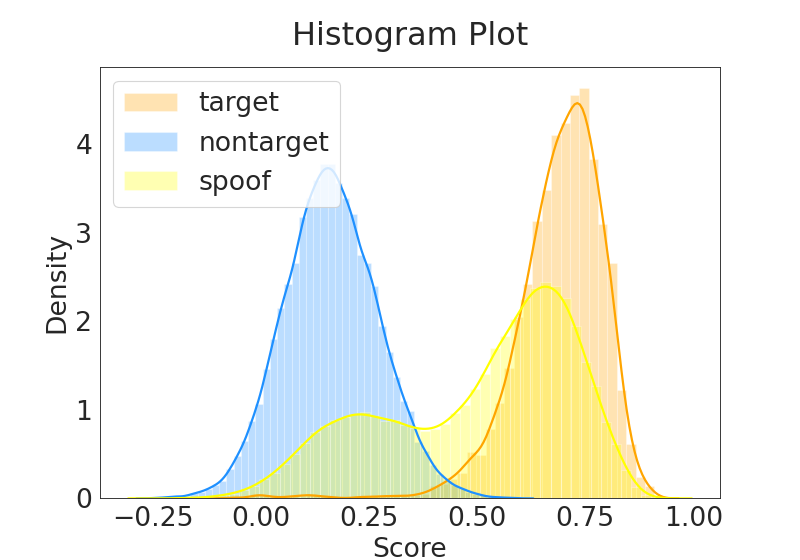}
\label{a.Conformer}
% \caption{a.Conformer}
}%
\subfigure[ECAPA-TDNN]{
\centering
\includegraphics[width=0.32\linewidth]{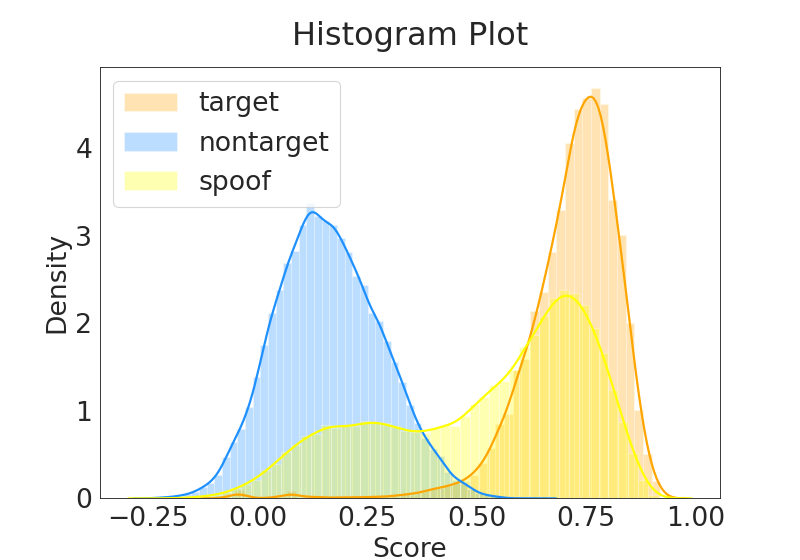}
\label{b.ECAPA-TDNN}
% \caption{b.fig2}
}% 
\subfigure[Resnet34]{
\centering
\includegraphics[width=0.32\linewidth]{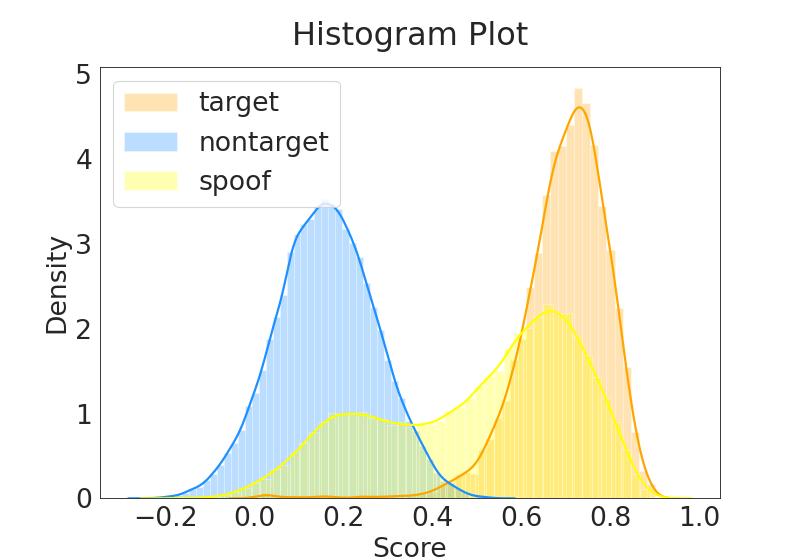}
\label{c.Resnet34}
% \caption{c.fig2}
}%
\quad
\subfigure[AASIST]{
\includegraphics[width=0.32\linewidth]{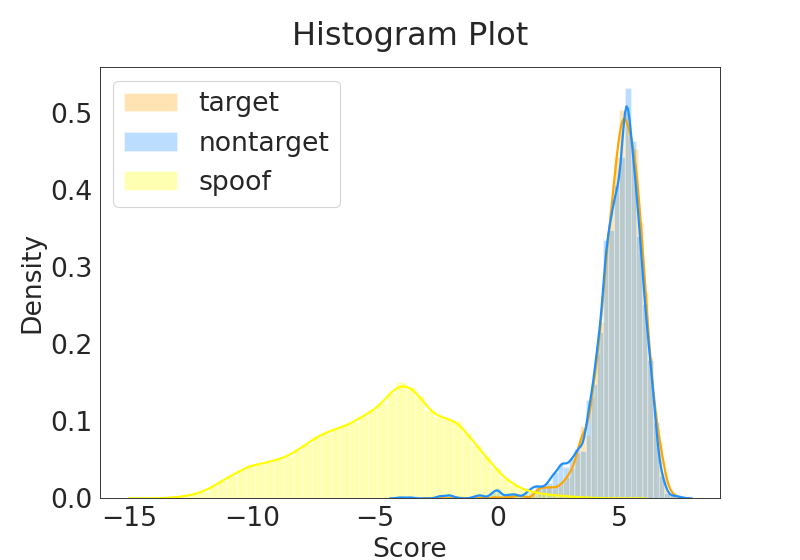}
%\caption{fig1}
}%
\subfigure[AASIST-L]{
\centering
\includegraphics[width=0.32\linewidth]{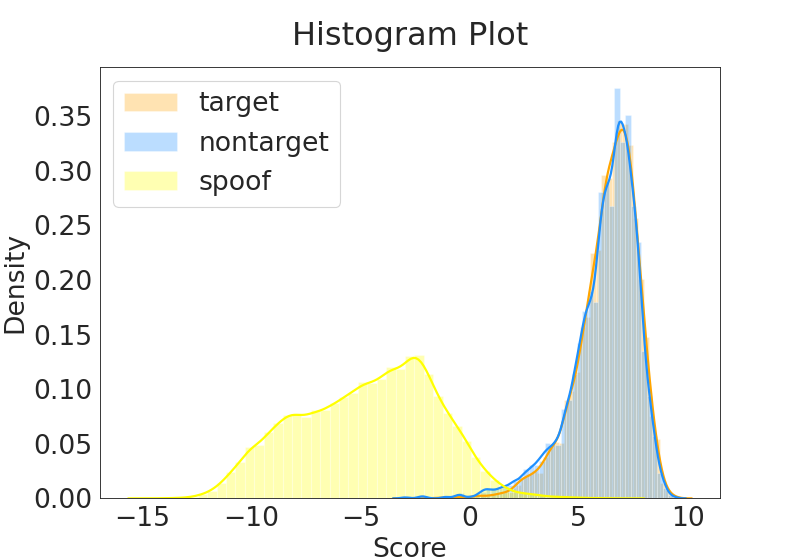}
%\caption{fig2}
}% 
\subfigure[RawGAT-ST]{
\centering
\includegraphics[width=0.32\linewidth]{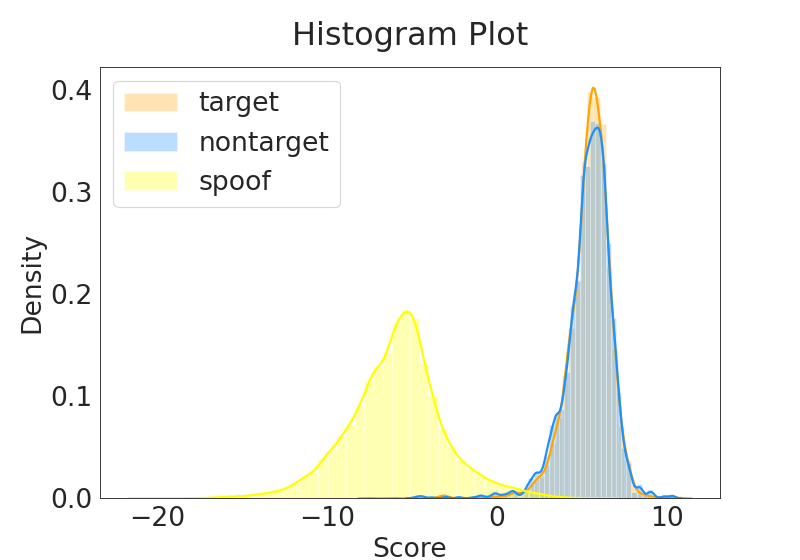}
%\caption{fig2}
}%
\quad
\subfigure[Baseline1]{
\includegraphics[width=0.32\linewidth]{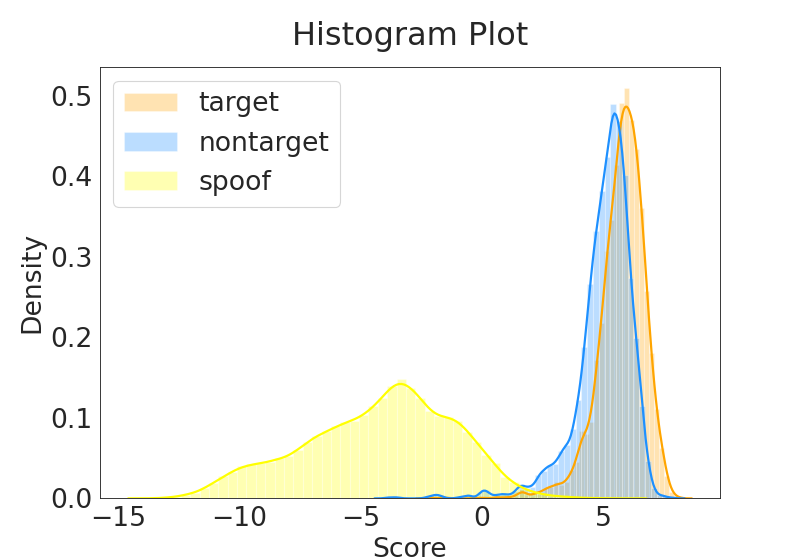}
%\caption{fig1}
}%
\subfigure[Baseline2]{
\centering
\includegraphics[width=0.32\linewidth]{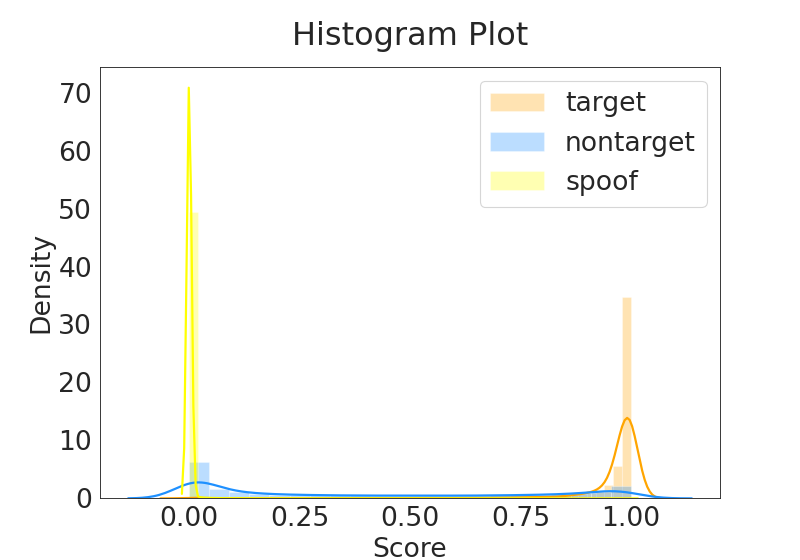}
%\caption{fig2}
}% 
\subfigure[SV-ALL\_CM-ALL]{
\centering
\includegraphics[width=0.32\linewidth]{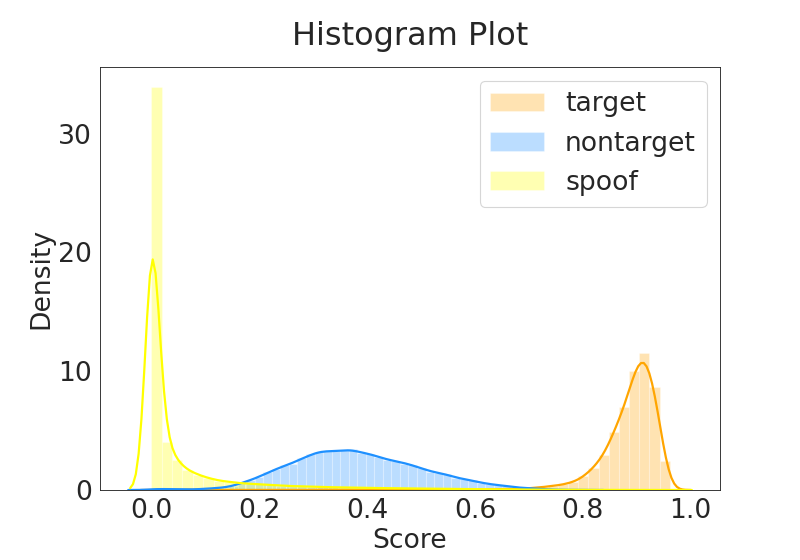}
%\caption{fig2}
}%
\centering
\caption{The histogram plots of different systems.}
\label{fig:hist}
\end{figure*}

\subsection{Experimental results}
Table~\ref{tab:all-eer} shows the three evaluation metrics provided by the SASV Challenge on the ASVspoof 2019 development and evaluation sets. A1-A3 denote using only speaker verification model. 
B1-B3 denote using only anti-spoofing model. 
C1 and C2 are the two baselines provided by the SASV challenge. 
D1-D9 denote using the fusion of one speaker verification model and one anti-spoofing model, e.g., MFA-Conformer\_AASIST means the combination of MFA-Conformer as speaker verification and AASIST as the anti-spoofing model. 
E1-E3 denote using the fusion of one speaker verification model with all the anti-spoofing models. 
F1-F3 denote the fusion of all three speaker verification models and one anti-spoofing model. 
G1 denotes the fusion of all the speaker verification models and anti-spoofing models. 

Figure~\ref{fig:hist} illustrates the histogram plots of all the systems.
All three kinds of trials, including target, non-target and spoof trials, are taken into account. 
The x-axis shows the prediction scores that indicate whether the trial is from the target or not. 
(a)-(c) are the plots for systems only using speaker verification models, including MFA-Conformer, ECAPA-TDNN and Resnet34. (d)-(f) are the plots for systems only using the anti-spoofing models, including AASIST, AASIST-L and RawGAT-ST. (g) and (h) are the plots for Baseline1 and Baseline2. (i) is the plot for the system using all three speaker verification models and anti-spoofing models for fusion.

\subsection{Observations and analysis}

We have the following observations and analysis:

\textit{Only using speaker verification models.} A1-A3 models, which are state-of-the-art ASV frameworks, performed well on the sub-task speaker verification, as shown in Table~\ref{tab:asv eer}.
Regarding the speaker verification performance, they generalize well to the ASVspoof 2019 evaluation set, e.g., the evaluation SV-EERs for the ECAPA-TDNN, MFA-Conformer and Resnet34 are 1.86\%, 1.38\% and 1.08\%, respectively.
However, they give poor results on detecting spoofing attacks, with evaluation SPF-EERs being 30.75\%, 30.22\% and 29.76\% for the ECAPA-TDNN, MFA-Conformer and Resnet34, respectively.
Also, as shown in Figure~\ref{fig:hist} (a)-(c), according to the distribution of the predicted scores, ASV models attain poor ability in distinguishing between spoofing trials and target trials, although they easily discriminate between target and non-target trials. 
Finally, the SASV-EERs are spoiled by the spoofing trials.
This phenomenon is reasonable, as the ASV is trained to distinguish different speakers.
And in order to acquire good speaker verification ability, the ASV model sometimes has to remove the incongruent information for speaker discrimination, such as microphone information, channel information, yet such information is helpful for anti-spoofing tasks.
    
\textit{Only using anti-spoofing models.} CM models (B1-B3) work well in detecting spoofing trials, showing satisfactory performance on SPF-EER, e.g. AASIST, AASIST-L and RawGAT-ST are with the evaluation SPF-EER of 0.67\%, 0.84\% and 0.96\% respectively as shown in Table~\ref{tab:all-eer}.
However, because of their speaker-unrelated training objective, they are incapable of verifying speakers, yielding SV-EERs close to 50\%, which indicates that their prediction for speaker verification task is nearly a random guess.
Unsurprisingly, we also note that the histograms of the target and non-target speakers almost completely overlap in Figure~\ref{fig:hist} (d)-(f), which is another evidence for their poor ability to distinguish among speakers.
As a result, the performance of spoofing-aware speaker verification of the countermeasure models are limited, e.g., AASIST, AASIST-L and RawGAT-ST are with the evaluation SASV-EERs of 24.38\%, 24.81\% and 24.85\%, respectively.
The anti-spoofing models are trained to distinguish whether an utterance is spoof or not, and the training criterion has no relation to speaker verification.
    
\textit{Two baselines.} Compared to the individual models, the two baseline fusion models are superior to using only ASV models or the spoofing countermeasure models alone in the SASV Challenge.
Among them, Baseline2 achieves an evaluation SASV-EER of 8.75\%, which is better than Baseline1 with an evaluation SASV-EER of 19.15\%, which is expectable.
This is because simple score-level fusion used in Baseline1 does not guarantee that the output scores of ASV and countermeasure belong to a unified space. 
The ASV scores range from -1 to 1, yet the countermeasure scores range from -20 to 15.
Computing simple additions will make the countermeasure score dominate the ASV score.
Figure~\ref{fig:hist} (g)(h) show that the distributions of the scores predicted by Baseline1 for target and non-target speak-ers are highly overlapped, while those of Baseline2 are much more separated.
% From Figure~\ref{fig:hist} (g)(h), the distributions of the scores predicted by Baseline1 for target and non-target speakers are highly overlapped, while those of Baseline2 are much more separable.
Baseline2 uses trainable deep neural networks to do better fusion among the ASV features and countermeasure features, and they can make a proportion of non-target trials distinguishable from the target trials as shown in Figure~\ref{fig:hist} (h).
However, the performance of Baseline2 on SV sub-task is still not satisfactory, which motivates us to explore further possibilities for fusion-based models.
    
\textit{Proposed approach.} The proposed models fuse 3 SOTA ASV models (A1-A3) and 3 SOTA anti-spoofing CM models (B1-B3). 
We have developed a total of 16 models with comprehensive experiments based on the proposed multi-model fusion framework. 
As shown in  Table~\ref{tab:all-eer}, all the proposed models outperform the baseline models with a large margin on the SASV-EER metrics, while retaining universally good performances on SV-EER and SPF-EER. 
% Furthermore, our best model G1, which is based on all the three ASV models and three anti-spoofing models, gets the evaluation SASV-EER of 1.17\% and outperform the Baseline2 with 7.58\% absolute (86.63\% relative). 
Empowered by the proposed framework, fusion models own significant discrimination capacity regarding the three tasks.
Among them, 9 models (3 by 3), D1-D9, are fused from the permutations of ASV-CM model pairs. D1-D3, which incorporate ASSIST as the CM model, have better results than fusion models incorporating other single CM model. 
D4-D6, which incorporate ASSIST-L as the CM model, show slight over-fitting on anti-spoofing task and lower EER on the ASV task compared with the respective individual models. 
D7-D9, which incorporate RawGAT-ST as the CM model, retain superior performance on anti-spoofing task but compromise the SV-EER. 
Further, we constructed 6 more models by involving all 3 ASV models with single CM model (E1-E3), and by involving single ASV model with all 3 CM models (F1-F3), respectively.
The models show better performance than D1-D9 in general, benefiting from the more comprehensive model ensemble. 
F2 shows over-fitting in the anti-spoofing and SASV tasks.
Finally, to push the limits of the multi-modal fusion framework, we incorporate all 3 ASV models and all 3 CM models and construct SV-ALL\_CM-A model (G1). 
As expected, the model yields a SASV-EER of 1.17\% given the evaluation trials, which gives the best performance among all models and achieves an 86.63\% relative boost compared to Baseline2.
Besides, G1 gave performances comparable with relevant SOTA models in the other two sub-tasks. 
Figure~\ref{fig:hist} (i) also shows that the distributions of the 3 classes are well-separated, which verifies the effectiveness of the proposed method.
% Besides, G1 owns comparable performances with relevant SOTA models regarding to other two sub-tasks. Figure~\ref{fig:hist} (i) also shows that the 3 classes' distributions are well-separated, verifying the effectiveness of the proposed method. 

%% file: 5-conclusion.tex
\section{Conclusion}
\label{sec:conclusion}

In order to take advantage of multiple state-of-the-art ASV and CM models, we propose a fusion-based framework to solve the SASV task. 
We comprehensively do the experiments to show the effectiveness of the proposed method.
The best fusion model effectively decreases the SASV-EER of the best baseline from 8.75\% to 1.17\%, achieving an improvement of 86\% relative (6.58\% absolute).

%% file: 6-acknowledge.tex
\section{Acknowledgements}
\label{sec:acknowledge}
This research is funded by the Centre for Perceptual and Interactive Intelligence, an InnoCentre of The Chinese University of Hong Kong.
This work was done while Haibin Wu was a visiting student at The Chinese University of Hong Kong.
Haibin Wu is supported by Google PhD Fellowship Scholarship.